\documentclass[aps,floatfix,prd,twocolumn]{revtex4}
\usepackage{graphicx}
\usepackage{dcolumn}
\usepackage{bm}
\usepackage{float}
\usepackage{amsmath}
\usepackage{csquotes}
\usepackage{romannum}
\usepackage{hyperref}
\hypersetup{
    colorlinks = true,
    citecolor = blue,
    linkcolor = blue,
    urlcolor = blue,
}

\begin{document}
\title{Influence of superconductivity on the magnetic moment of quantum impurity embedded in BCS superconductor}
\author{Sachin Verma and Ajay}
\affiliation{
Department of Physics,
Indian Institute of Technology,
Roorkee, Uttarakhand 247667,
INDIA.\\
\text{\bf{Email:}} sverma2@ph.iitr.ac.in and ajay@ph.iitr.ac.in}

\begin{abstract}
We study the influence of superconductivity on the formation of the localized magnetic moment for a single-level quantum impurity embedded in an s-wave Bardeen-Cooper-Schrieffer (BCS) superconducting medium, modeled by single-impurity Anderson Hamiltonian. We have combined Bogoliubov transformation with Green's function method within self-consistent Hartree-Fock Mean Field approximation to analyze the conditions necessary in metal (in the superconducting) for the formation of the magnetic moment at the impurity site for the low-frequency limit $|\omega|<<\Delta_{sc}$ as well as for the finite superconducting gap $\Delta_{sc}$. We have compared these results with other theoretical results and with the single-level quantum impurity embedded in the normal metallic host. Further, we analyze the spectral density of the quantum impurity embedded in a superconducting host to study the sub-gap states as a function of impurity parameters.\\
\\
\textbf{\textit{KEYWORDS --- quantum impurity, superconductivity, magnetic moment, Anderson model, Hartree-Fock Mean-field approximation}}
\end{abstract}
\maketitle
\section{Introduction}
Advancement in nano-fabrication techniques made it possible to make devices in which metallic leads are connected to quantum dots (QDs)\cite{Kouwenhoven2001, Franceschi2010,Martin2011}. In principle, such hybrid devices are experimental realization for quantum impurities interacting with the sea of conduction band electrons. The quantum dots are nanoscopic semiconductor structures (e.g., InAs nanowire, carbon nanotube, and Graphene QDs) in which electrons are confined to zero dimensions. Due to the quantum confinement, these quantum dots have a discrete energy level like an atom. These hybrid combinations of metal and QDs are used as a single electron transistor(SET), a device that is highly conductive only at very specific gate voltage. If the metal is in the superconducting state, then it provides vital applications, such as nano-SQUID for the detection of the individual magnetic molecule, sources of spin entangled electrons, and detectors for mechanical resonators\cite{Franceschi2010}. Quantum dot connected to one or more superconducting leads can be used to study the exciting phenomenon, e.g., quantum phase transition or interplay between Kondo physics and superconductivity\\
In his pioneering work, Anderson\cite{Anderson1961} analyzed the electronic structure of a metal containing a quantum impurity and studied the conditions necessary in metals for the presence or absence of localized magnetic moments at the impurity site. By using self-consistent Hartree-Fock approximation (HFA), it has been shown that local magnetic moments might be formed under suitable conditions determined by an interplay of certain physical parameters such as impurity energy levels, the coupling between impurity and metal (hybridization energy or s-d interaction) and on-site Coulomb repulsion.\\
The quantum dot (impurity) embedded in a superconducting bath has also been a topic of intensive research from the past few decades. Quantum impurity embedded in a superconducting bath (with superconducting energy gap $\Delta_{sc}$) instead of a normal metal drastically modifies the electronic structure of impurity. The proximity effect allows the Cooper pair to leak into the quantum impurity state, thus shows an induced pairing in its spectral function in the energy range $-\Delta_{sc}<\omega<\Delta_{sc}$. Andreev reflection (i.e., conversion of electrons into the Cooper pairs with a simultaneous reflection of the holes) on the opposite interfaces give rise to discrete sub-gap states known as Andreev bound states (ABSs). The presence of strong on-site Coulomb interaction opposes any double occupancy of the quantum impurity state (Coulomb blockade). At low temperatures, the magnetic impurity can be screened by the conduction electrons at the Fermi sea(i.e., the formation of a spin S=0 state by breaking the Cooper pairs)\cite{Kondo1964,Glazman2001}. This so-called Kondo effect occurs if the impurity is strongly coupled to the bath(i.e., for small $\Delta_{sc}/\Gamma$, where $\Gamma$ is the coupling between impurity and superconducting bath), which then hybridizes with the impurity level. Thus both the effect, i.e., the on-site Coulomb repulsion and appearance of Kondo singlet competes with the superconductivity.\\
The effect of superconductivity on the formation of localized magnetic moments has been studied previously within the Hartree-Fock approximation by various authors \cite{Tripathi1967,Kusakabe1971,Rossler1972}. Early studies showed that superconductivity hinders the spin localization or magnetic moment formation in comparison with normal metals by assuming that there is no pairing induced by superconductivity on the impurity site\cite{Tripathi1967}.
 Kusakabe \cite{Kusakabe1971} and Rossler and Kiwi \cite{Rossler1972} included the pairing induced on the localized site of the impurity. Kusakabe concluded that the superconductivity either aided or hindered the formation of the localized magnetic moment according to the impurity level's energy relative to the Fermi surface. On the other hand, Rossler and Kiwi find out that the magnetic region is slightly reduced relative to the normal state.
More recently, the spectroscopic properties of quantum impurity embedded in a superconducting host are studied in the superconducting atomic limit ($\Delta_{sc}\rightarrow\infty$) \cite{Bauer2007,Baranski2013,Domanski2008,Vecino2003,Meng2009}. Such a limiting case provides a useful way to handle the problem analytically.\\
 Previous Hartree-Fock treatment of superconductor quantum dot nanostructures were either incomplete \cite{Rozhkov1999,Zhu2001} (i.e exclude self-consistent determination of the induced gap), computationally incorrect \cite{Yoshioka2000}, more focused on the effect of magnetic impurity on the superconductivity \cite{Shiba1973} and study the formation of the sub-gap Andreev bound states and transport properties of superconductor quantum dot Josephson junction \cite{Martin2012}. The recent experimental and theoretical study on the effect of one or more magnetic impurity on the bulk superconductor is given in references \cite{Balatsky2006,Meng2015,Heinrich2018}. Further the transport properties of superconductor quantum dot Josephson junction have also been extensively analyzed by using second-order perturbation theory \cite{Vecino2003,Meng2009,Zonda2015,Zonda2016} and renormalization group techniques \cite{Choi2004,Karrasch2008,Lim2008,Wentzell2016}. The experimental and theoretical studies of superconductor quantum dot nanostructures have been summarized in Ref.\cite{Franceschi2010,Martin2011,Meden2019}.\\
\begin{figure}[h]
  \includegraphics[scale=0.45]{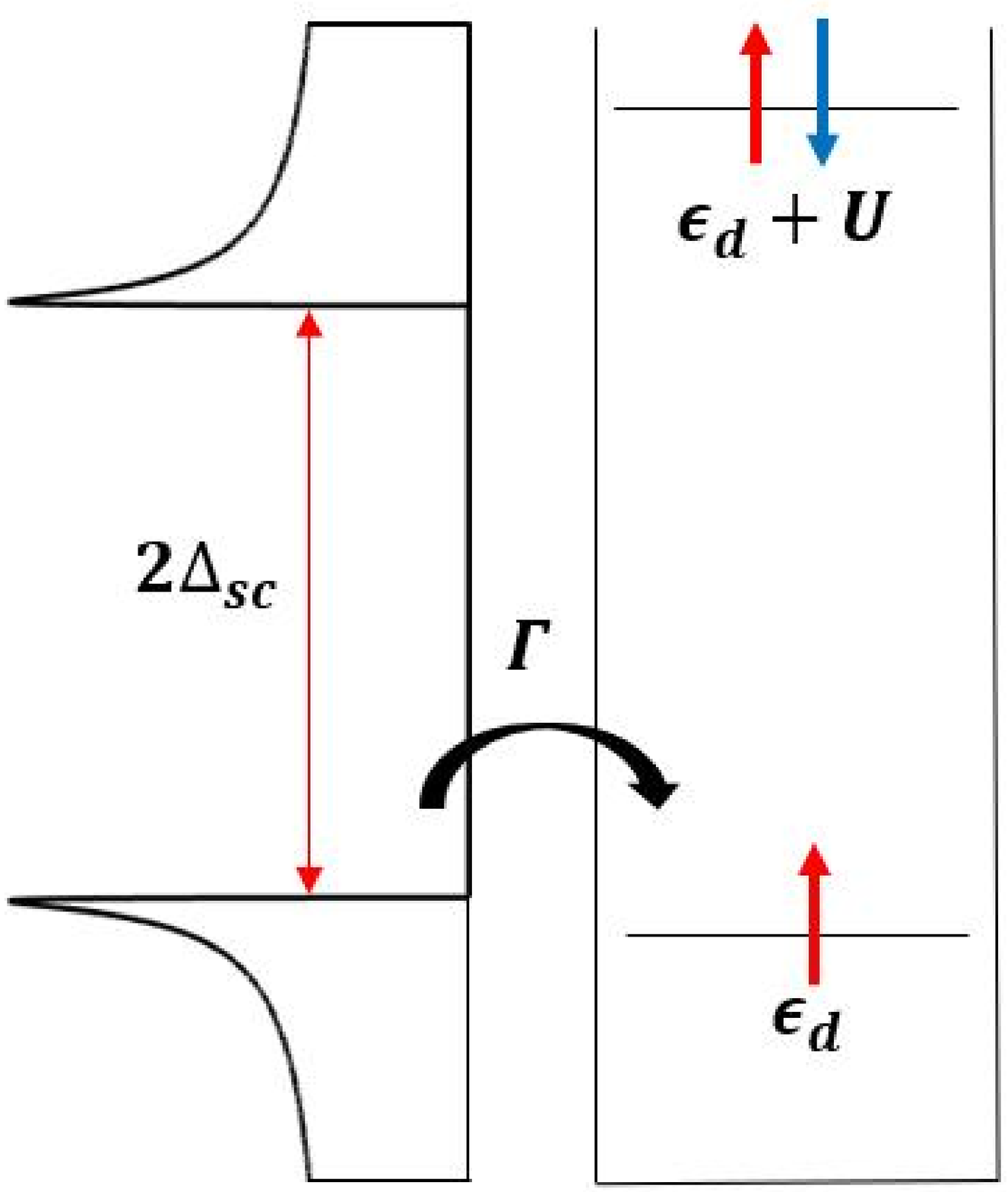}
 \caption {Schematic level diagram for quantum impurity (QD) embedded in BCS superconductor: an impurity level $\epsilon_d$ with on-site Coulomb repulsion $U$ is hybridized with a continuum of excitations in a superconductor with a gap $\Delta_{sc}$ through hybridization with strength $\Gamma$.}
\end{figure}\\
The tunneling spectroscopy experiment of Andreev bound states in carbon nanotube quantum dot or InAs nanowire quantum dot contacted with normal and/or superconducting leads has been studied in detail\cite{Deacon2010,Lee2014,Pillet2013,Buitelaar2002,vanDam2006,Maurand2012,Li2017}. Tunnel spectroscopy experiment by Deacon et al \cite{Deacon2010} provides the evidence of singlet (non-magnetic) to doublet (magnetic) transition in S-QD-N (with negligible coupling to the normal lead) when the number of electrons changes from even to odd. Further, Lee et al \cite{Lee2014} studied the magnetic properties of S-QD. The experimental study of the electronic and transport properties of the hybrid superconductor quantum dot Josephson Junction (S-QD-S) can be found in Ref. \cite{Pillet2013,Buitelaar2002,vanDam2006,Maurand2012,Li2017}. Maurand et al \cite{Maurand2012} study a carbon-nanotube quantum dot embedded in a superconducting quantum interference device to investigate the competition of strong coulomb correlation with induced pairing. In the strong Coulomb blockade regime, the singlet to doublet transition ($0-\pi$ transition for Josephson junctions) is controlled by a change in the energy of the quantum dot level relative to the Fermi level. At a larger coupling, the Kondo effect develops for a magnetic impurity(dot), and suppress magnetism. The competition between the singlet and doublet states is governed by different energy scales: superconducting gap ($\Delta_{sc}$), the coupling between the superconductor and quantum dot ($\Gamma$), the on-site Coulomb correlation or charging energy ($U$), and the energy ($\epsilon_d$) of the dot level relative to the Fermi energy of the Superconducting electrode (see Fig 1)\\
Motivated by the above theoretical and experimental studies, we have planned to investigate the single impurity Anderson model with a superconducting bath within self-consistent HFA. This approximation gives qualitative insight into the lowest order correlation effect but does not capture the Kondo physics. We focused on the sub-gap states and the formation and stability of finite magnetic moment at the quantum impurity site immersed in an s-wave superconducting host for the low-frequency limit $|\omega|<<\Delta_{sc}$(i.e., a gap much larger than all characteristic impurity energies) and for the finite superconducting gap case. It is also assumed that the energy level spacing of quantum impurity $\delta\epsilon$ is sufficiently large compared to other energy parameters. Thus the impurity is simplified to a two-fold degeneracy of spin up and spin down (by Pauli exclusion principle).\\
Kondo effect can also arise in impurity coupled to superconductor in the low-frequency limit if there exists a Fermionic state near Fermi level \cite{Baranski2013}. For example, in N-QD-S junctions, itinerant electrons from the normal electrode can screen magnetic impurity giving rise to the Kondo effect. For $|\omega|<<\Delta_{sc}$, the single quantum impurity is coupled only to a superconductor (S-QD). Thus there are no single-particle states near the Fermi level, and as a result, the formation of the Kondo peak is suppressed. The coupling between quantum impurity and quasiparticle excitations also vanishes in the low-frequency limit. But the impurity is still coupled to the Cooper pairs at the Fermi level, which induces a superconducting gap (proportional to the hybridization $\Gamma$ between impurity and superconductor) in the spectral density of impurity. Thus in the low-frequency limit, the on-site Coulomb repulsion competes with the induced on dot pairing. For the finite superconducting gap $\Delta_{sc}$, the remnants of the Kondo effect can influence the singlet to doublet transition by suppressing the magnetism \cite{Maurand2012,Maurand2013}. However, the general picture of singlet to doublet transition in non-Kondo regime ($T_K<<\Delta_{sc}$, where $T_K=\sqrt{\frac{U\Gamma}{2}}e^{\left(\frac{-\pi U}{8\Gamma}\right)}$) is captured on mean-field level. To describe the non-magnetic singlet to magnetic doublet transition in the Kondo regime ($T_K>>\Delta_{sc}$), it is, however, necessary to go beyond the mean-field approximation, taking into account the higher-order dynamical correlations. A detailed discussion of the model Hamiltonian and theoretical formulation is provided in the preceding section \Romannum{2}.
\section{Model Hamiltonian and Theoretical formulation}
Single-level Anderson impurity Model provides the microscopic model Hamiltonian for a single-level quantum impurity (QD) embedded in BCS superconducting bath,
\begin{equation}\label{eq:pareto mle2}
  \begin{aligned}
\hat{H}=\hat{H}_{QD}+\hat{H}_S+\hat{H}_T
 \end{aligned}
\end{equation}
where
\begin{gather} 
\begin{aligned}
\hat{H}_{QD}=\sum_{\sigma} \epsilon_{d}\hat{n}_{d\sigma}+U\hat{n}_{d\uparrow}\hat{n}_{d\downarrow}
 \end{aligned} \\
  \begin{aligned}
\hat{H}_S=\sum_{k,\sigma}(\epsilon_{k}\hat{c}^\dagger_{k\sigma}\hat{c}_{k\sigma})-\sum_{k}(\Delta_{sc}\hat{c}^\dagger_{k\uparrow}\hat{c}^\dagger_{-k\downarrow}+\Delta^\ast_{sc}\hat{c}_{-k\downarrow}\hat{c}_{k\uparrow})
\end{aligned} \\
\begin{aligned}
\hat{H}_T=\sum_{k,\sigma}(V_{k}\hat{d}^\dagger_{\sigma}\hat{c}_{k\sigma}+{V^\ast_{k}}\hat{c}^\dagger_{k\sigma}\hat{d}_{\sigma})
 \end{aligned} 
 \end{gather}
$\hat{H}_{QD}$ (Eq.(2)) is the Hamiltonian for single-level quantum impurity, $d_\sigma(d^\dagger_\sigma)$ is the annihilation(creation) operator of electron with spin $\sigma$ on the impurity and $n_{d\sigma}=d_\sigma^\dagger d_\sigma$ is number operator. The impurity consists of a single electronic level of energy $\epsilon_d$ and can be occupied upto two electrons. The Coulomb repulsion $U$ between electrons on the impurity state is also taken into account, which hinders an exact solution to the problem.\\
 $\hat{H}_S$ (Eq.(3)) is BCS Hamiltonian, $c_{k\sigma}(c^\dagger_{k\sigma})$ is the annihilation(creation) operator of an electron with spin $\sigma$ and wave vector $\vec{k}$ in the superconducting bath.
 In $\hat{H}_S$, the first term is the kinetic energy, and the second term represents the attractive interaction between the electrons of the superconducting bath, which is responsible for the formation of Cooper pairs. $\Delta_{sc}$ is a superconducting energy gap i.e., energy difference between the ground state of the superconductor and energy of lowest quasiparticle excitations.
 The energy $\epsilon_k$ is measured with respect to the chemical potential $\mu_S=\epsilon_f=0$ at $T=0K$.\\
$\hat{H}_T$ (Eq.(4)) represents the hybridization of impurity with the external superconducting bath, i.e., the possibility of the single-particle tunneling between impurity state and superconducting bath and vice-versa. $V_{k}$ is the hybridization energy (or s-d interaction).\\
To diagonalized the BCS part of the above Hamiltonian, we employ the so-called Bogoliubov transformation. We define new Fermionic quasiparticle operators $\gamma_{k\sigma}$ and coefficients $u_k$ and $v_k$ 
\begin{equation}\label{eq:pareto mle2}
\begin{aligned}
c_{k\uparrow} = u^\ast_k\gamma_{k\uparrow}+v_k\gamma^\dagger_{-k\downarrow}
\\
c^\dagger_{-k\downarrow} = u_k\gamma^\dagger_{-k\downarrow}-v^\ast_k\gamma_{k\uparrow}
\end{aligned}
\end{equation}
The normalization condition is $|{u_k}|^2+|{v_k}|^2=1$. 
Substituting in the Eq.(1) yields 
\begin{equation}\label{eq:pareto mle2}
  \begin{aligned}
H=\sum_{k,\sigma}(E_{k}\gamma^\dagger_{k\sigma}\gamma_{k\sigma}+E_0)
+\sum_{k\sigma}(V_ku^\ast_kd^\dagger_{\sigma}\gamma_{k\sigma}+h.c)+
\\
\sum_{k}[V^\ast_kv_k(d^\dagger_{\uparrow}\gamma^\dagger_{-k\downarrow}-d^\dagger_{\downarrow}\gamma^\dagger_{k\uparrow})+ h.c]
+\sum_{\sigma} \epsilon_{d}n_{d\sigma}+Un_{d\uparrow}n_{d\downarrow}
 \end{aligned}
\end{equation}
where $h.c$ denotes the Hermitian conjugate, $E_0=\sum_k(\epsilon_k-E_k+\Delta_{sc}\langle{c^\dagger_{k\uparrow}c^\dagger_{-k\downarrow}}\rangle)$ is the ground state energy of the bath and  $E_k = \sqrt{\epsilon^2_k+|\Delta_{sc}|^2}$ is the excitation energy (quasiparticle energy)  of the bath.
  We assume that hybridization or s-d interaction is $k$ independent i.e $V_k=V$ for $V_k<<D$ (wide band) where $-D\leq\epsilon_k\leq D$ and the normal tunneling rate from dot to the lead (or coupling constant) is defined by $\Gamma=\pi|V|^2\rho_0$, where normal density of states $\rho_0$ is constant in the range of energies around the Fermi level (flat band).\\
The coefficients $u_k$ and $v_k$ read
\begin{gather}
|{u_k}|^2=\frac{1}{2}(1+\frac{\epsilon_k}{\sqrt{{\epsilon_k}^2+|\Delta_{sc}|^2}})
\\
 |{v_k}|^2=\frac{1}{2}(1-\frac{\epsilon_k}{\sqrt{{\epsilon_k}^2+|\Delta_{sc}|^2}})
  \end{gather} \\
For $\Delta_{sc}\rightarrow 0$, $|{u_k}|^2\rightarrow 1$ for $\epsilon_k>0$ and $|{u_k}|^2\rightarrow 0$ for  $\epsilon_k<0$ whereas ${|v_k}|^2\rightarrow 1$ for $\epsilon_k<0$ and ${|v_k}|^2\rightarrow 0$ for  $\epsilon_k>0$. Thus a Bogoliubon excitation in normal state corresponds to creating an electron for energies above the Fermi level and creating a hole of opposite momentum and spin for energies below the Fermi level. At the superconducting state, a Bogoliubon becomes a superposition of both electron and hole state.\\
 To solve the above single level Anderson impurity model, we use the Green's function Equation of motion (EOM) method \cite{Zubarev1960}
We are mainly interested in the spectral properties of the quantum impurity which can be extracted from the single-particle retarded Green's function defined as
\begin{equation}\label{eq:pareto mle2}
\begin{aligned}
G^r_{d\sigma}(t) = \langle\langle{d_{\sigma}(t);d^\dagger_{\sigma}(0)}\rangle\rangle=-i\theta(t)\langle{[d_{\sigma}(t),d^\dagger_{\sigma}(0)]_+}\rangle
\end{aligned}
\end{equation}
In the framework of the Green's function method, the Fourier transform of the single particle retarded Green function should satisfy the equation of motion
\begin{equation}\label{eq:pareto mle2}
\begin{aligned}
\omega \langle\langle{d_{\sigma};d^\dagger_{\sigma}}\rangle\rangle_\omega=\langle{[d_{\sigma},d^\dagger_{\sigma}]_+}\rangle+\langle\langle{[d_{\sigma},H];d^\dagger_{\sigma}}\rangle\rangle_\omega
\end{aligned}
\end{equation}
 For correlated ($U\neq0$) quantum impurity embedded in a superconducting bath, the Hamiltonian is not exactly solvable due to the quartic term in the Coulomb interaction. Therefore we analyze above Hamiltonian by treating the Coulomb interaction within HFA.\\
 For the finite superconducting gap $\Delta_{sc}$, the interaction term in HFA is written as \cite{Shiba1973, Yoshioka2000, Martin2012, Vecino2003} 
 \begin{gather}
 \begin{aligned}
U{n}_{d\uparrow}{n}_{d\downarrow}=U\langle{n}_{d\uparrow}\rangle{n}_{d\downarrow}+U\langle{n}_{d\downarrow}\rangle{n}_{d\uparrow}+ \\
U\langle{d^\dagger_{\downarrow} d^\dagger_{\uparrow}}\rangle d_{\uparrow} d_{\downarrow}+ U\langle{d_{\uparrow} d_{\downarrow}}\rangle d^\dagger_{\downarrow} d^\dagger_{\uparrow}
\end{aligned}
\end{gather}
where $\langle{n}_{d\sigma}\rangle$ is the average number of occupation of spin $\sigma\in\;\uparrow,\downarrow$ on the dot and the dimensionless parameter $\langle{d_{\uparrow} d_{\downarrow}}\rangle$ or $\langle{d^\dagger_{\downarrow} d^\dagger_{\uparrow}}\rangle$ are the qualitative measure of the induced on-dot pairing.\\
Thus the Hamiltonian (Eq.(6)) becomes
\begin{gather}
\begin{aligned}
H={} & \sum_{k,\sigma}(E_{k}\gamma^\dagger_{k\sigma}\gamma_{k\sigma}+E_0)+\sum_{k\sigma}(V_ku^\ast_kd^\dagger_{\sigma}\gamma_{k\sigma}+h.c) + \\
& \sum_{k}[V^\ast_kv_k(d^\dagger_{\uparrow}\gamma^\dagger_{-k\downarrow}-d^\dagger_{\downarrow}\gamma^\dagger_{k\uparrow})+ h.c] +  \\
& \sum_{\sigma}E_{d\sigma}d^\dagger_{\sigma}d_{\sigma}+ U\langle{d_{\uparrow} d_{\downarrow}}\rangle \left( d_{\uparrow} d_{\downarrow} + d^\dagger_{\downarrow} d^\dagger_{\uparrow}\right)
\end{aligned}                                                
\end{gather}
where $E_{d\sigma}=\epsilon_{d}+U\langle{n}_{d\bar{\sigma}}\rangle$ and $\sigma\neq\bar{\sigma}$\\
Both  $\langle{n}_{d\sigma}\rangle$ and $\langle{d_{\uparrow} d_{\downarrow}}\rangle$ have to be calculated self-consistently.
Again, by employing the Green's function EOM method, we derived the following coupled equations for the case of correlated quantum impurity.
\begin{gather}
\begin{aligned}
(\omega-E_{d\uparrow})\langle\langle{d_{\uparrow};d^\dagger_{\uparrow}}\rangle\rangle=1+V\sum_ku^\ast_k\langle\langle{\gamma_{k\uparrow};d^\dagger_{\uparrow}}\rangle\rangle+\\
V\sum_kv_k\langle\langle{\gamma^\dagger_{-k\downarrow};d^\dagger_{\uparrow}}\rangle\rangle-U\langle{d_{\uparrow} d_{\downarrow}}\rangle \langle\langle{d^\dagger_{\downarrow}; d^\dagger_{\uparrow}}\rangle\rangle
\end{aligned}                                                
\end{gather}
\begin{gather}
\begin{aligned}
(\omega-E_k)\langle\langle{\gamma_{k\uparrow};d^\dagger_{\uparrow}}\rangle\rangle=V^\ast u_k\langle\langle{d_{\uparrow};d^\dagger_{\uparrow}}\rangle\rangle+\\
V v_k\langle\langle{d^\dagger_{\downarrow};d^\dagger_{\uparrow}}\rangle\rangle
\end{aligned}                                                
\end{gather}
\begin{gather}
\begin{aligned}
(\omega+E_k)\langle\langle{\gamma^\dagger_{-k\downarrow};d^\dagger_{\uparrow}}\rangle\rangle=-V u^\ast_k\langle\langle{d^\dagger_{\downarrow};d^\dagger_{\uparrow}}\rangle\rangle+\\
V^\ast v^\ast_k\langle\langle{d_{\uparrow};d^\dagger_{\uparrow}}\rangle\rangle
\end{aligned}                                                
\end{gather}
and 
\begin{gather}
\begin{aligned}
(\omega+E_{d\downarrow})\langle\langle{d^\dagger_{\downarrow};d^\dagger_{\uparrow}}\rangle\rangle=-V^\ast\sum_ku_k\langle\langle{\gamma^\dagger_{-k\downarrow};d^\dagger_{\uparrow}}\rangle\rangle+\\
V^\ast\sum_kv^\ast_k\langle\langle{\gamma_{k\uparrow};d^\dagger_{\uparrow}}\rangle\rangle-U\langle{d_{\uparrow} d_{\downarrow}}\rangle \langle\langle{d_{\uparrow};d^\dagger_{\uparrow}}\rangle\rangle
\end{aligned}                                                
\end{gather}
We solved above close set of equations (Eqs.13-16) to find the expression for the single electron retarded Green's function $(\langle\langle{d_{\uparrow};d^\dagger_{\uparrow}}\rangle\rangle)$ of the quantum dot.
\begin{equation}
\resizebox{1.0\hsize}{!}{$G^r_{d}(\omega)=\frac{\omega+E_{d\downarrow}-I_1}{(\omega+E_{d\downarrow}-I_1)(\omega-E_{d\uparrow}-I_2)-(I_3+ U\langle{d_{\uparrow} d_{\downarrow}}\rangle)^2}$}
\end{equation}
where
\begin{equation}
I_1=|V|^2\sum_k\left(\frac{|u_k|^2}{\omega+E_k}+\frac{|v_k|^2}{\omega-E_k}\right)
\end{equation}
\begin{equation}
I_2=|V|^2\sum_k\left(\frac{|u_k|^2}{\omega-E_k}+\frac{|v_k|^2}{\omega+E_k}\right)
\end{equation}
and
\begin{equation}
I_3=|V|^2\sum_ku_kv^\ast_k\left(\frac{1}{\omega+E_k}-\frac{1}{\omega-E_k}\right)
\end{equation}
Where $I_1$ and $I_2$ are the diagonal and $I_3$ is the off-diagonal part of selfenergy (which corresponds to induced paring) due to coupling between QD and superconducting host in Nambu representation \cite{Bauer2007,Baranski2013}.
By transferring summation over $k$-values into the integral over $\epsilon$, the multi-dimensional problem changes into a one-dimensional problem and for $|\omega|<\Delta_{sc}$ i.e. within superconducting gap one can have
$$I_1=I_2=-2|V|^2\rho_0\omega\int_{0}^{D\rightarrow\infty} \left[\frac{1}{\epsilon^2+\left(\Delta^2_{sc}-\omega^2\right)}\right]d\epsilon$$
\begin{equation}
=-\frac{\Gamma\omega}{\sqrt{\Delta^2_{sc}-\omega^2}}
\end{equation}
and
$$I_3=2|V|^2\rho_0\Delta_{sc}\int_{0}^{\infty} \left[\frac{1}{\epsilon^2+\left(\Delta^2_{sc}-\omega^2\right)}\right]d\epsilon$$
\begin{equation}
=\frac{\Gamma\Delta_{sc}}{\sqrt{\Delta^2_{sc}-\omega^2}}
\end{equation}
And for $|\omega|>\Delta{sc}$ a simple manipulation provides
\begin{equation}
I_1=I_2=-\frac{i\Gamma\omega}{\sqrt{\Delta_{sc}^2-\omega^2}}
\end{equation}

\begin{equation}
I_3=\frac{i\Gamma\Delta_{sc}}{\sqrt{\Delta_{sc}^2-\omega^2}}
\end{equation}
In the next subsections, we calculate single electron retarded Green's function and corresponding spectral density for uncorrelated quantum impurity and correlated quantum impurity  for the finite superconducting gap and low-frequency limit $|\omega|<<\Delta_{sc}$.
\subsection{Non-interacting case($U=0$)}
For the non-interacting case or uncorrelated quantum impurity $E_{d\sigma}=\epsilon_d$, and the Hamiltonian is exactly solvable and the Green function of the quantum dot is given by (from Eq. 17),
\begin{gather} 
G^r_{d0}(\omega)=\frac{\omega+\epsilon_d+\frac{\Gamma\omega}{\sqrt{\Delta^2_{sc}-\omega^2}}}{\omega^2+\frac{2\Gamma\omega^2}{\sqrt{\Delta^2_{sc}-\omega^2}}-\epsilon^2_d-\Gamma^2} \;  \; , \; \; |\omega|<\Delta_{sc} 
\end{gather}
\begin{gather} 
G^r_{d0}(\omega)=\frac{\omega+\epsilon_d+\frac{i\Gamma\omega}{\sqrt{\omega^2-\Delta^2_{sc}}}}{\omega^2+\frac{2i\Gamma\omega^2}{\sqrt{\omega^2-\Delta^2_{sc}}}-\epsilon^2_d-\Gamma^2} \;  \; , \; \; |\omega|>\Delta_{sc}
\end{gather}
For $|\omega|<\Delta_{sc}$ the poles of Green's function (i.e. equating denominator to zero) gives the energies of localized excited states or Andreev bound states (ABSs) which can be obtained by solving following equation
\begin{gather} 
\omega^2-\left[\frac{\left(\epsilon^2_d +\Gamma^2\right)}{\left(1+\frac{2\Gamma}{\sqrt{\Delta^2_{sc}-\omega^2}}\right)}\right]=0
\end{gather}
The Green's function for $|\omega|<\Delta_{sc}$ can also be written as ($\omega\rightarrow\omega+i\delta$)
\begin{gather} 
G^r_{d0}(\omega+i\delta)=\sum_{\alpha=\pm}{\frac{W^{\alpha}_b}{\omega-E^{\alpha}_b+i\delta}}
\end{gather}
where $E^{+}_b =E_b$ and $E^{-}_b =-E_b$ are the poles of Green function i.e solution of Eq.(27) while their respective spectral weights  i.e., $W^+_b$ and  $W^-_b$ are calculated from their residuals
\begin{equation}
W^{\alpha}_b = \left[\frac{\omega+\epsilon_d+\frac{\Gamma\omega}{\sqrt{\omega^2-\Delta^2_{sc}}}} {\frac{d}{d\omega}\left(\omega^2+\frac{\Gamma\omega^2}{\sqrt{\omega^2-\Delta^2_{sc}}}-\epsilon^2_d-\Gamma^2\right)}\right]_{\omega=E^{\alpha}_b}
\end{equation}\\
\begin{equation}
\resizebox{1.0\hsize}{!}{$W^{\alpha}_b = \frac{1}{2}\left(\Delta^2_{sc}-E^2_b\right)\left[\frac{\sqrt{\Delta^2_{sc}-E^2_b}(1+\frac{\alpha\epsilon_d}{E_b})+\Gamma}{(\Delta^2_{sc}-E^2_b)(\sqrt{\Delta^2_{sc}-E^2_b}+2\Gamma)+\Gamma E^2_b}\right]$}
\end{equation}
For uncorrelated QD in the particle-hole symmetric case ($\epsilon_d=-U/2$), the weights of ABSs become equal i.e $W^{+}_b=W^{-}_b$.\\
The parameter $\delta\rightarrow 0^+$ is equivalent to the infinitesimally weak coupling to the normal lead in S-QD-N system \cite{Baranski2013, Domanski2008}.  Weak coupling to the normal lead  means that it only serves as a probe to provide the information of quantum dot, without disturbing the quantum states there. Finite coupling to the normal lead changes the width of the sub-gap states i.e the lifetime of quasiparticles ($\delta\propto\frac{1}{\tau}$). Thus, in our analysis of the S-QD system ($\delta=10^{-5}\Gamma$) the sub-gap states become Dirac delta function, i.e. they represent the quasiparticles of an infinite life-time.\\
The total spectral density on the quantum impurity is defined as the imaginary part of the retarded Green’s function $\rho_{d0}=-\frac{1}{\pi}Im\{G^r_{d0}(\omega)\}$ (here only considering spin up; the total spectral density is
the sum of spin up and spin down spectral density, which are equal by symmetry).
 \begin{gather}
 \rho_{d0}(\omega)= \frac{\delta}{\pi}\sum_{\alpha=\pm}\left[\frac{W^{\alpha}_b}{(\omega-E^{\alpha}_b)^2+\delta^2}\right]+\rho_{cont}(\omega)
 \end{gather}
 with 

\begin{equation}
\resizebox{1.0\hsize}{!}{$\rho_{cont}(\omega)=\frac{1}{\pi}\frac{\Gamma\omega}{\sqrt{\omega^2-\Delta^2_{sc}}}\left[\frac{\omega^2+2\epsilon_d\omega+\Gamma^2+\epsilon^2_d}{\omega^2-(\Gamma^2+\epsilon^2_d)^2+\left(\frac{2\Gamma\omega^2}{\sqrt{\omega^2-\Delta^2_{sc}}}\right)^2}\right]$}
\end{equation}\\
 where first term is the discrete spectral density for $|\omega|<\Delta_{sc}$ and second term $\rho_{cont}(\omega)$ is continuum spectral density for $|\omega|>\Delta_{sc}$.
 \subsection{Interacting case($U\neq0)$ : For finite Superconducting gap ($\Delta_{sc}$)}
For interacting or correlated quantum impurity, the single-particle retarded Green's functions (Eq.(17)) and the anomalous Green's function, which corresponding to the pairing parameter, is calculated by using Eqs.(13)-(16) above \cite{Bauer2007,Martin2012,Wentzell2016}.
\begin{equation}
\resizebox{1.0\hsize}{!}{$G^r_{d,11}(\omega)=\langle\langle{d_{\uparrow};d^\dagger_{\uparrow}}\rangle\rangle=\left[\frac{\left(\omega+E_{d\downarrow}+\frac{\Gamma\omega}{\sqrt{\Delta^2_{sc}-\omega^2}}\right)}{\left(\omega+E_{d\downarrow}+\frac{\Gamma\omega}{\sqrt{\Delta^2_{sc}-\omega^2}}\right)\left(\omega-E_{d\uparrow}+\frac{\Gamma\omega}{\sqrt{\Delta^2_{sc}-\omega^2}}\right)-\left(\frac{\Gamma\Delta_{sc}}{\sqrt{\Delta^2_{sc}-\omega^2}}+U\langle{d_{\uparrow} d_{\downarrow}}\rangle\right)^2}\right]$}
\end{equation}
where $\omega\rightarrow\omega+i\delta$
\begin{equation}
\resizebox{1.0\hsize}{!}{$G^r_{d,22}(\omega)=\langle\langle{d^\dagger_{\downarrow};d_{\downarrow}}\rangle\rangle=\left[\frac{\left(\omega-E_{d\downarrow}+\frac{\Gamma\omega}{\sqrt{\Delta^2_{sc}-\omega^2}}\right)}{\left(\omega-E_{d\downarrow}+\frac{\Gamma\omega}{\sqrt{\Delta^2_{sc}-\omega^2}}\right)\left(\omega+E_{d\uparrow}+\frac{\Gamma\omega}{\sqrt{\Delta^2_{sc}-\omega^2}}\right)-\left(\frac{\Gamma\Delta_{sc}}{\sqrt{\Delta^2_{sc}-\omega^2}}+U\langle{d_{\uparrow} d_{\downarrow}}\right)^2}\right]$}
\end{equation}
\begin{equation}
\resizebox{1.0\hsize}{!}{$G^r_{d,21}(\omega)=\langle\langle{d^\dagger_{\downarrow};d^\dagger_{\uparrow}}\rangle\rangle=\left[\frac{\left(\frac{\Gamma\Delta_{sc}}{\sqrt{\Delta^2_{sc}-\omega^2}}+U\langle{d_{\uparrow} d_{\downarrow}}\right)}{\left(\omega+E_{d\downarrow}+\frac{\Gamma\omega}{\sqrt{\Delta^2_{sc}-\omega^2}}\right)\left(\omega-E_{d\uparrow}+\frac{\Gamma\omega}{\sqrt{\Delta^2_{sc}-\omega^2}}\right)-\left(\frac{\Gamma\Delta_{sc}}{\sqrt{\Delta^2_{sc}-\omega^2}}+U\langle{d_{\uparrow} d_{\downarrow}}\right)^2}\right]$}
\end{equation}
The self-consistent equation for the average occupation number  $\langle{n_{d\sigma}}\rangle$ at the quantum dot level of a given spin $\sigma$ and the pairing parameter $\langle{d_{\uparrow} d_{\downarrow}}\rangle$ is obtained by integrating corresponding Spectral density in the continuum energy up to the Fermi level $\epsilon_f$ (at $T=0K$).
\begin{gather}
 \langle{n_{d\sigma}}\rangle=-\frac{1}{\pi}\int_{-\infty}^0 Im\{\langle\langle{d_{\sigma};d^\dagger_{\sigma}}\rangle\rangle \} d\omega
 \end{gather}
 \begin{gather}
 \langle{d_{\uparrow} d_{\downarrow}}\rangle=\langle{d^\dagger_{\downarrow};d^\dagger_{\uparrow}}\rangle=-\frac{1}{\pi}\int_{-\infty}^0 Im\{\langle\langle{d^\dagger_{\downarrow};d^\dagger_{\uparrow}}\rangle\rangle \} d\omega
 \end{gather}
 Starting from the initial guess for occupation $\langle n_{d\uparrow}\rangle_{(1)} = 0.5$, we iterate three self-consistent Hartree-Fock equations for  $\langle n_{d\uparrow}\rangle$,  $\langle n_{d\downarrow}\rangle$ and  $\langle{d_{\uparrow} d_{\downarrow}}\rangle$, $k$ times until $\langle n_{d\uparrow}\rangle_{(k+1)}-\langle n_{d\uparrow}\rangle_{(k)} \leq 10^{-6}$.\\
 The spectral density of the quantum dot features discrete ABS inside the superconducting gap. Thus  the spectral density of quantum dot is given by\cite{Wentzell2016}
 \begin{equation}
 \rho_d(\omega)=-\frac{1}{\pi}Im\left[G^r_{d,11}(\omega)+G^r_{d,22}(\omega)\right]
 \end{equation}
 \subsection{Interacting case($U\neq0)$ : For low frequency limit ($|\omega|<<\Delta_{sc}$)}
  For correlated quantum impurity, the simple solvable limit is the limit of large gap i.e., $\Delta_{sc}\rightarrow\infty$, and has been discussed previously \cite{Bauer2007,Baranski2013,Domanski2008,Vecino2003,Meng2009}. This is not the limit, as realized in the real experiment, but it allows one to obtain the exact analytical solution, and it becomes more useful for the complex multi-terminal and/or multi-dot nanostructures.
 We  take the low-frequency limit $|\omega|<<\Delta_{sc}$ after taking $D\rightarrow\infty$ for the proximity effect to survive which is equivalent to $\Delta_{sc}\rightarrow\infty$ limit.\\
The Hamiltonian for Correlated quantum impurity coupled to BCS superconductor (Eq.(6)) is
\begin{gather}
\begin{aligned}
H={} & \sum_{k,\sigma}(E_{k}\gamma^\dagger_{k\sigma}\gamma_{k\sigma}+E_0)+\sum_{k\sigma}(V u^\ast_kd^\dagger_{\sigma}\gamma_{k\sigma}+h.c)+ \\
& \sum_{k}[V^\ast v_k(d^\dagger_{\uparrow}\gamma^\dagger_{-k\downarrow}-d^\dagger_{\downarrow}\gamma^\dagger_{k\uparrow})+ h.c]+\sum_{\sigma}E_{d\sigma}d^\dagger_{\sigma}d_{\sigma}
\end{aligned}                                                
\end{gather}
where $E_{d\sigma}=\epsilon_{d}+U\langle{n}_{d\bar{\sigma}}\rangle$ and $\sigma\neq\bar{\sigma}$\\
By employing the Green's function EOM method, we derived the following coupled equations for the case of the correlated quantum dot.
\begin{gather}
\begin{aligned}
(\omega-E_{d\uparrow})\langle\langle{d_{\uparrow};d^\dagger_{\uparrow}}\rangle\rangle=1+V\sum_ku^\ast_k\langle\langle{\gamma_{k\uparrow};d^\dagger_{\uparrow}}\rangle\rangle+\\
V\sum_kv_k\langle\langle{\gamma^\dagger_{-k\downarrow};d^\dagger_{\uparrow}}\rangle\rangle
\end{aligned}                                                
\end{gather}
\begin{gather}
\begin{aligned}
(\omega-E_k)\langle\langle{\gamma_{k\uparrow};d^\dagger_{\uparrow}}\rangle\rangle=V^\ast u_k\langle\langle{d_{\uparrow};d^\dagger_{\uparrow}}\rangle\rangle+\\
V v_k\langle\langle{d^\dagger_{\downarrow};d^\dagger_{\uparrow}}\rangle\rangle
\end{aligned}                                                
\end{gather}
\begin{gather}
\begin{aligned}
(\omega+E_k)\langle\langle{\gamma^\dagger_{-k\downarrow};d^\dagger_{\uparrow}}\rangle\rangle=-V u^\ast_k\langle\langle{d^\dagger_{\downarrow};d^\dagger_{\uparrow}}\rangle\rangle+\\
V^\ast v^\ast_k\langle\langle{d_{\uparrow};d^\dagger_{\uparrow}}\rangle\rangle
\end{aligned}                                                
\end{gather}
and 
\begin{gather}
\begin{aligned}
(\omega+E_{d\downarrow})\langle\langle{d^\dagger_{\downarrow};d^\dagger_{\uparrow}}\rangle\rangle=-V^\ast\sum_ku_k\langle\langle{\gamma^\dagger_{-k\downarrow};d^\dagger_{\uparrow}}\rangle\rangle+\\
V^\ast\sum_kv^\ast_k\langle\langle{\gamma_{k\uparrow};d^\dagger_{\uparrow}}\rangle\rangle
\end{aligned}                                                
\end{gather}
Then we can solve this close set of equations (Eqs.40-43) and get the expression for the single electron retarded Green's function  as follows
\begin{equation}
\resizebox{1.0\hsize}{!}{$G^r_{d}(\omega)=\langle\langle{d_{\uparrow};d^\dagger_{\uparrow}}\rangle\rangle=\frac{\omega+E_{d\downarrow}-I_1}{(\omega-E_{d\uparrow}-I_2)(\omega+E_{d\downarrow}-I_1)-(I_3)^2}$}
\end{equation}
Where for $|\omega|<\Delta_{sc}$ we have
\begin{equation}
I_1=-\frac{\Gamma\omega}{\sqrt{\Delta^2_{sc}-\omega^2}}
\end{equation}
and
\begin{equation}
I_3=\frac{\Gamma\Delta_{sc}}{\sqrt{\Delta^2_{sc}-\omega^2}}
\end{equation}
In the low-frequency regime $|\omega|<<\Delta_{sc}$ Eqs. (45) and (46)become
$$I_1=I_2=0, I_3=\Gamma.$$
Thus the Green function of correlated quantum (Eq.(44)) in the above limit become
\begin{equation}
G^r_{d}(\omega)=\frac{\omega+E_{d\downarrow}}{(\omega-E_{d{\uparrow}})(\omega+E_{d\downarrow})-\left(\Gamma\right)^2}
\end{equation}
where $\omega\rightarrow\omega+i\delta$\\
The energies of ABSs is given by the poles of the Green's function i.e 
$$D(\omega)=(\omega-E_{d\bar{\sigma}})(\omega+E_{d\sigma})-\left(\Gamma\right)^2=0$$
  The corresponding spectral density and average occupation number for spin $\sigma$ on the quantum dot is given by,
 \begin{gather}
 \rho_{d\sigma}(\omega)=-\frac{1}{\pi}Im\{G^r_{d\sigma}(\omega)\} 
 \end{gather}
 \begin{equation}
 \langle n_{d\sigma}\rangle=\left[\frac{1}{2}-\frac{E_{d\sigma}}{\pi\sqrt{E^2_{d\sigma}+\Delta^2_d}} \tan^{-1}\left(\frac{\sqrt{E^2_{d\sigma}+\Delta^2_d}}{\delta}\right)\right]
 \end{equation}
 where $E_{d\sigma}=\epsilon_d+U\langle n_{d\bar{\sigma}}\rangle$ and $\Delta_d = \Gamma$ is proximity induced superconducting gap i.e for low frequency limit $\omega<<\Delta_{sc}$ quantum dot itself become superconducting grain with induced gap equal to $\Gamma$ (similar to the uncorrelated quantum dot with $\Gamma<<\Delta_{sc}$).\\
The above equation's form is the same as obtained by \cite{Anderson1961} for quantum impurity embedded in a normal metallic host.
These two coupled equations for $\langle n_{d\uparrow}\rangle$ and $\langle n_{d\downarrow}\rangle$ give us the occupation of the spin up and spin down states at the impurity site and the magnetic moment, $m=\langle n_{d\uparrow}\rangle-\langle n_{d\downarrow}\rangle$.\\
 In the next section, we discuss the results obtained by the numerical computations for various parameter regimes.
\section{Results and Conclusion}
The competition between the non-magnetic (singlet) and magnetic (doublet) ground state  at the impurity site embedded in superconductor host is determined by different energy scales: $\Delta_{sc}$, $\Gamma$, $U$ and $\epsilon_d$.\\
In Fig.2 we present the spectral density $\rho_{d0}(\omega)$ (Eq.(31)) for different coupling strengths $\Gamma$ for uncorrelated ($U=0$) quantum impurity. This shows that the spectral density $\rho_{d0}(\omega)$ vanishes for $\omega<\Delta_{sc}$ i.e inside the superconducting gap except for certain discrete values. These resonant sub-gap states (ABSs) represents the quasiparticles of infinite lifetime ($\tau=1/\delta$) and is the signature of the proximity effect. 
\begin{figure}[h]
\includegraphics[scale=0.28]{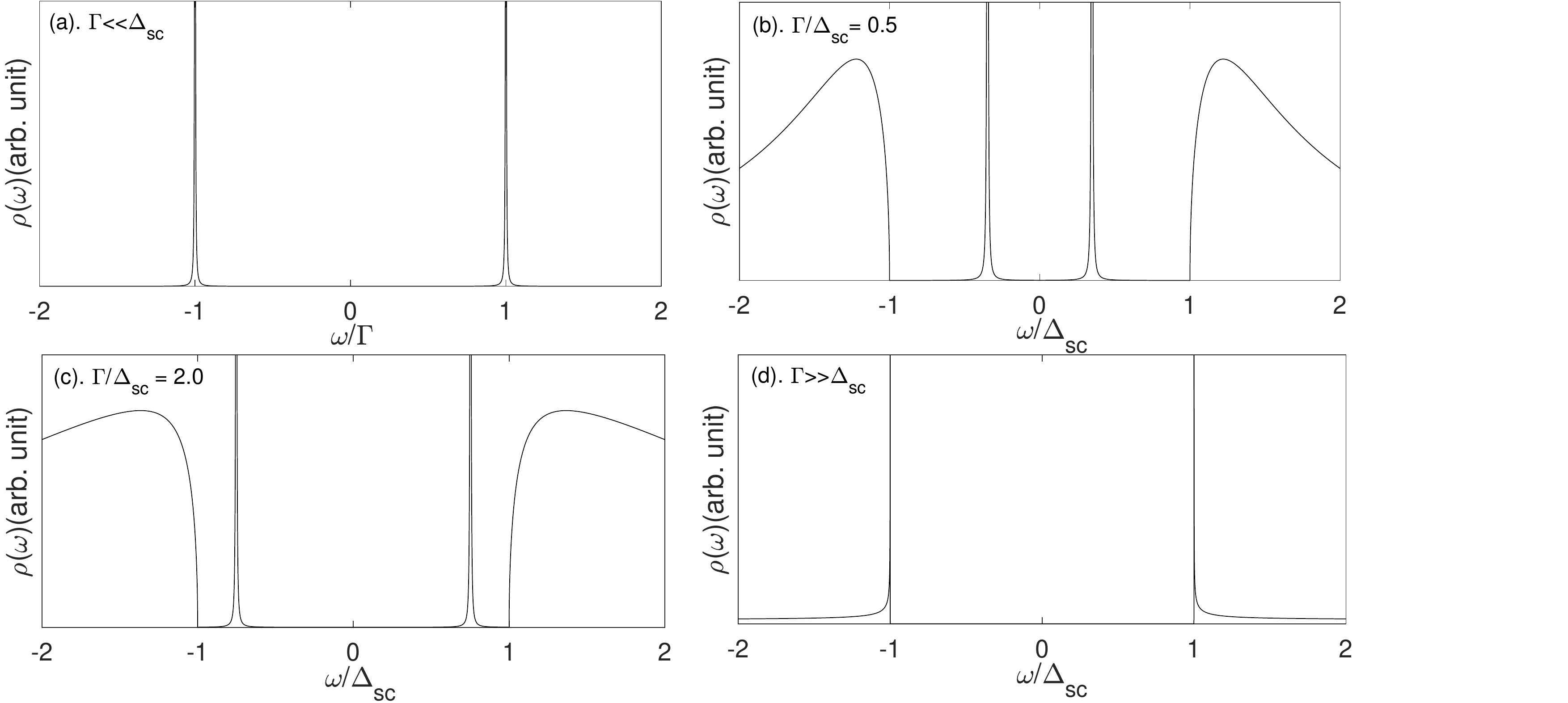}
\centering
  \caption{The spectral density $\rho_{d0}(\omega)$ of uncorrelated quantum impurity obtained for various value of $\Gamma/\Delta_{sc}$ for electron-hole symmetric case ($\epsilon_d=-U/2=0$) and $\delta\rightarrow 0^+$ at $ T=0K$.}
\label{fig:nonfloat}
\end{figure}\\
In the superconducting atomic limit($\Gamma<<\Delta_{sc}$ or $\Delta_{sc}\rightarrow\infty$) the poles of Green's function is $\omega=\pm\sqrt{\epsilon^2_d+\Gamma^2}$, which shows that in this limit the impurity itself become superconductor with the induced pairing gap $\Delta_d=|\Gamma|$ as shown in Fig 2.(a). This superconducting singlet $|S\rangle$ ($S=0$) is superposition of the empty,$ |0\rangle$, and doubly occupied, $|\uparrow\downarrow\rangle$,states i.e $|S\rangle= -v^{\ast}_d|\uparrow\downarrow\rangle+u_d|0\rangle$.
On the other hand, for $\Gamma>>\Delta_{sc}$, i.e., in a strong coupling regime, the resonant sub-gap quasiparticles states combine with the gap edge singularities at $\pm\Delta{sc}$ as shown in Fig 2.(d). Figure 2.(b) and 2.(c) shows the well defined sub-gap ABSs for intermediate values of $\Gamma/\Delta_{sc}$.\\
In the low-frequency regime $|\omega|<<\Delta_{sc}$, the proximity induced on-dot pairing $|\Delta_d|=\Gamma$ competes with Coulomb repulsion $U$. For $U=0$ case, it follows from Eqs.(49) that $$\langle n_{d\uparrow}\rangle=\langle n_{d\downarrow}\rangle$$ 
which is a non-magnetic solution as discussed above.
\begin{figure}[h]
  \includegraphics[scale=0.26]{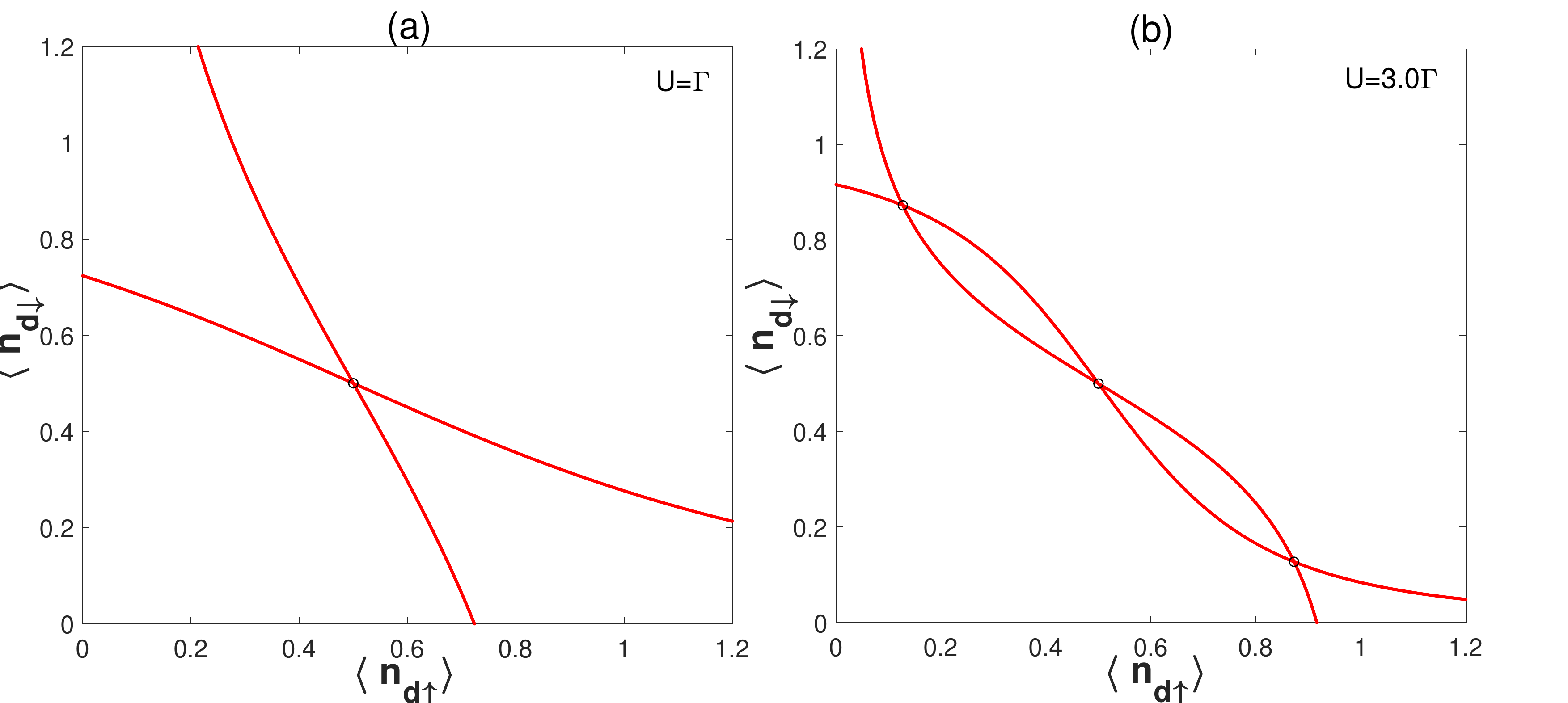}
  \caption{Self -consistent plot of $\langle n_{d\uparrow}\rangle$ vs $\langle n_{d\downarrow}\rangle$ for a) Non-Magnetic case  and b) Magnetic case for electron-hole symmetry ($\epsilon_d=\frac{-U}{2}$).}
  \label{fig:nonfloat}
\end{figure}\\
The self-consistent solution of Eqs.(49) for $\langle n_{d\uparrow}\rangle$ and $\langle n_{d\downarrow}\rangle$ shows the existence of the singlet ($|S\rangle$) and magnetic doublet ( $|\uparrow\rangle$, $|\downarrow\rangle$ ) solution for different values of $U/\Gamma$ (see Fig.3 and Fig.4).
For small $U$ i.e  $U/\Gamma=1.0$, there exists only one non-magnetic solution at $\langle n_{d\uparrow}\rangle$=$\langle n_{d\downarrow}\rangle=1/2$. But for $U/\Gamma=3.0$ we find the \enquote{localized} case with three possible solutions, one non-magnetic solution at $\langle n_{d\uparrow}\rangle$=$\langle n_{d\downarrow}\rangle=1/2$ and another pair of stable magnetic solution at $\langle n_{d\uparrow}\rangle$=$1-\langle n_{d\downarrow}\rangle=0.87266$. The existence of this spurious  spin-symmetry breaking ($\langle n_{d\uparrow}\rangle\neq\langle n_{d\downarrow}\rangle$) or spontaneous generation of Zeeman field is not present in the exact solution. Despite this spurious symmetry breaking, the HFA contains the physics of magnetization due to Coulomb correlation and  provides qualitatively fine description of singlet to doublet transition and the corresponding phase diagram for weak ($\Gamma<<\Delta_{sc}$)and intermediate ($\Gamma\leq\Delta_{sc}$) coupling strength with small on site coulomb interaction $U$.
\begin{figure}[h]
  \includegraphics[scale=0.27]{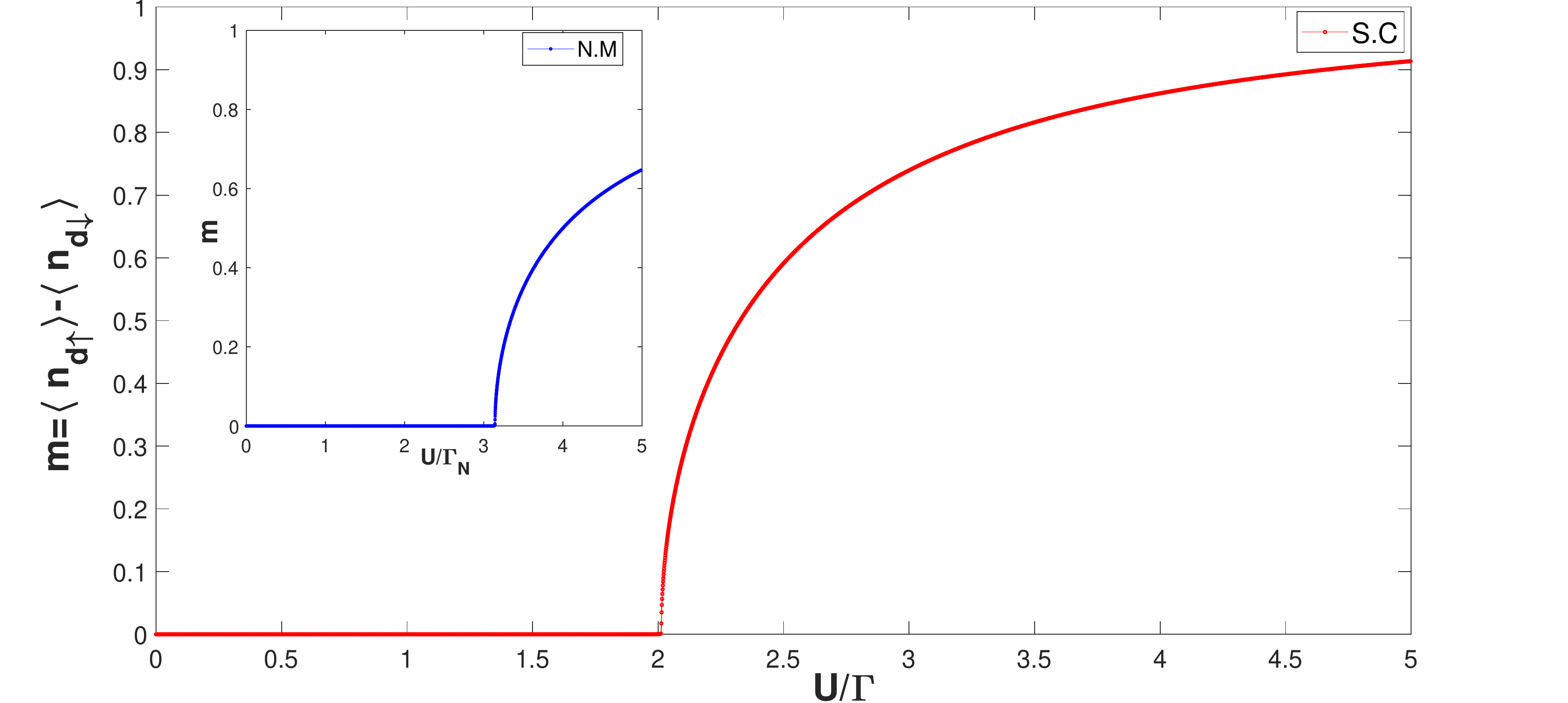}
  \centering
  \caption{$U/\Gamma$ dependence of the magnetic moment $m$ at the impurity site embedded in S.C for $\epsilon_d=-U/2$ at $T=0K$ in the low-frequency regime $|\omega|<<\Delta_{sc}$. The inset shows the $U/\Gamma_N$ dependence of the magnetic moment $m$ at the impurity site embedded in Normal metal (N.M) host.}
  \label{fig:nonfloat}
\end{figure}\\
Fig.4 shows the  $U/\Gamma$ dependence of magnetic moment $m=\langle n_{d\uparrow}\rangle-\langle n_{d\downarrow}\rangle$ at the impurity site for electron-hole symmetric case i.e  $\epsilon_d=-U/2$. The ground state is singlet as long as $U\leq 2\Gamma$ and doublet otherwise. Thus the ground state transition occurs at  $U/2\Gamma\simeq 1$. In the superconducting case, the Andreev bound state within the superconducting gap causes the singlet to doublet transition. For a quantum impurity embedded in a normal metallic host (N-QD) with tunnel coupling $\Gamma_N$, HFA leads to the magnetic phase transition at $U/\pi\Gamma_N=1$ (see inset of Fig.4). Thus singlet to doublet transition in N-QD (in non-Kondo regime) occurs for a larger value of on-site Coulomb interaction $U$ as compared to S-QD ($\Delta_{sc}\rightarrow\infty$).\\
Let us now discuss the magnetic moment as a function of the energy level of quantum impurity, $\epsilon_d$, away from the electron-hole symmetric case ($\epsilon_d\neq -U/2$). The phase diagram depicts the stability of the magnetic doublet versus that of the spin-singlet.\\
 Fig.5(b) is the phase diagram for impurity embedded in the superconducting host, and it shows the magnetic moment in color-scale representation as a function of $\epsilon_d/U$ and $\Gamma/U$. This Singlet to  Doublet transition is consistent with the previous results (exact for particle-hole symmetric case) \cite{Bauer2007,Meng2009}. These authors considered the effective localized model with $\Delta\rightarrow\infty$ to study sub-gap states and to obtain the phase boundary analytically. The equation of phase boundary is given by $\Gamma/U=\sqrt{1/4-(\epsilon_d/U+1/2)^2}$ (black dashed line in figure 5(b)).\\
Fig.5(a) shows the phase diagram for impurity embedded in normal metal \cite{Anderson1961}. If the metal is in the superconducting state with $|\omega|<<\Delta_{sc}$ then the magnetic region is enhanced as compared to the normal metallic state by a factor of 1.9 (i.e. area of a magnetic doublet in a superconducting case $\approx$ 1.9 $\times$ area of a magnetic doublet in normal case). The  enhancement of the magnetic region is due to the change in the density of states near the Fermi level of the host metal.\\
\begin{figure}[h]
  \includegraphics[scale=0.26]{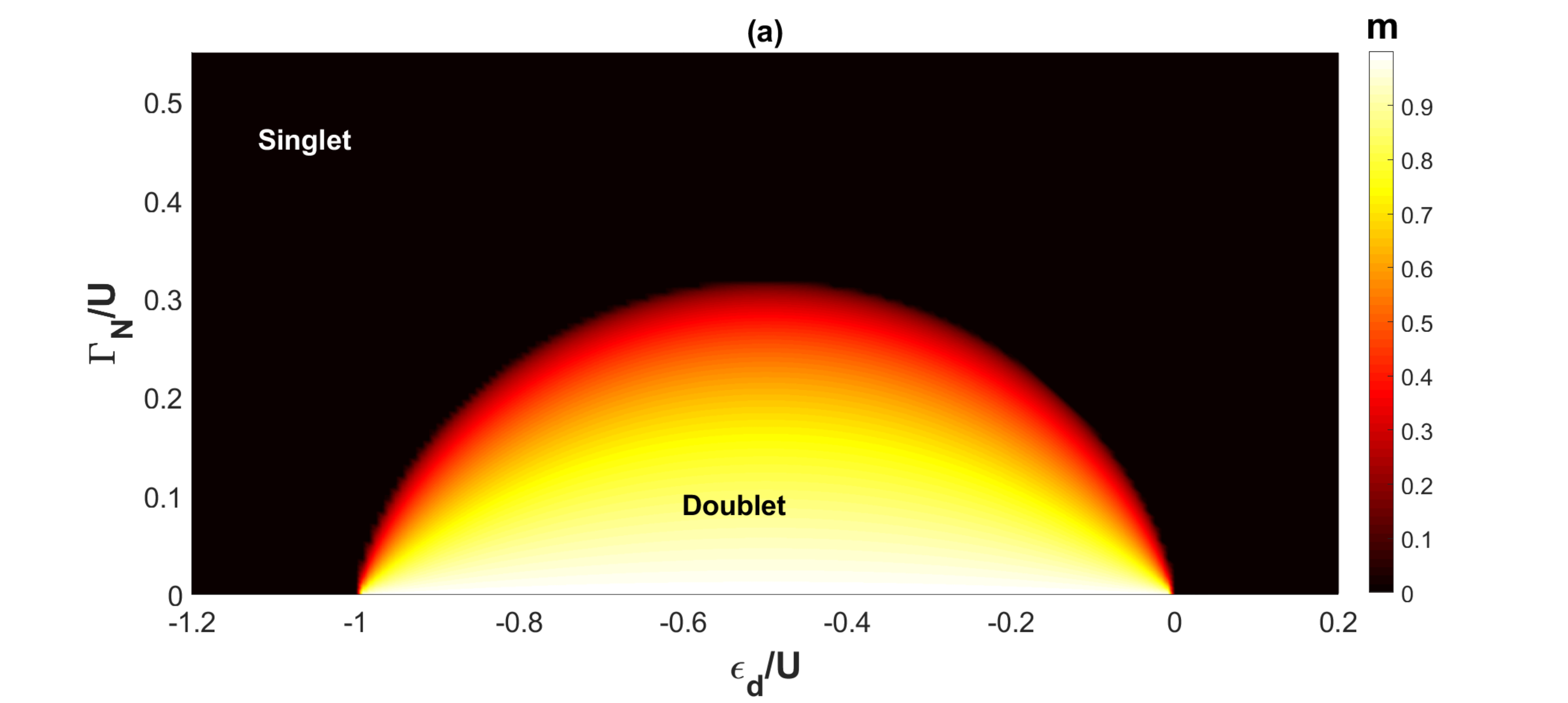}
  \centering
\end{figure}
\begin{figure}[h]
  \includegraphics[scale=0.26]{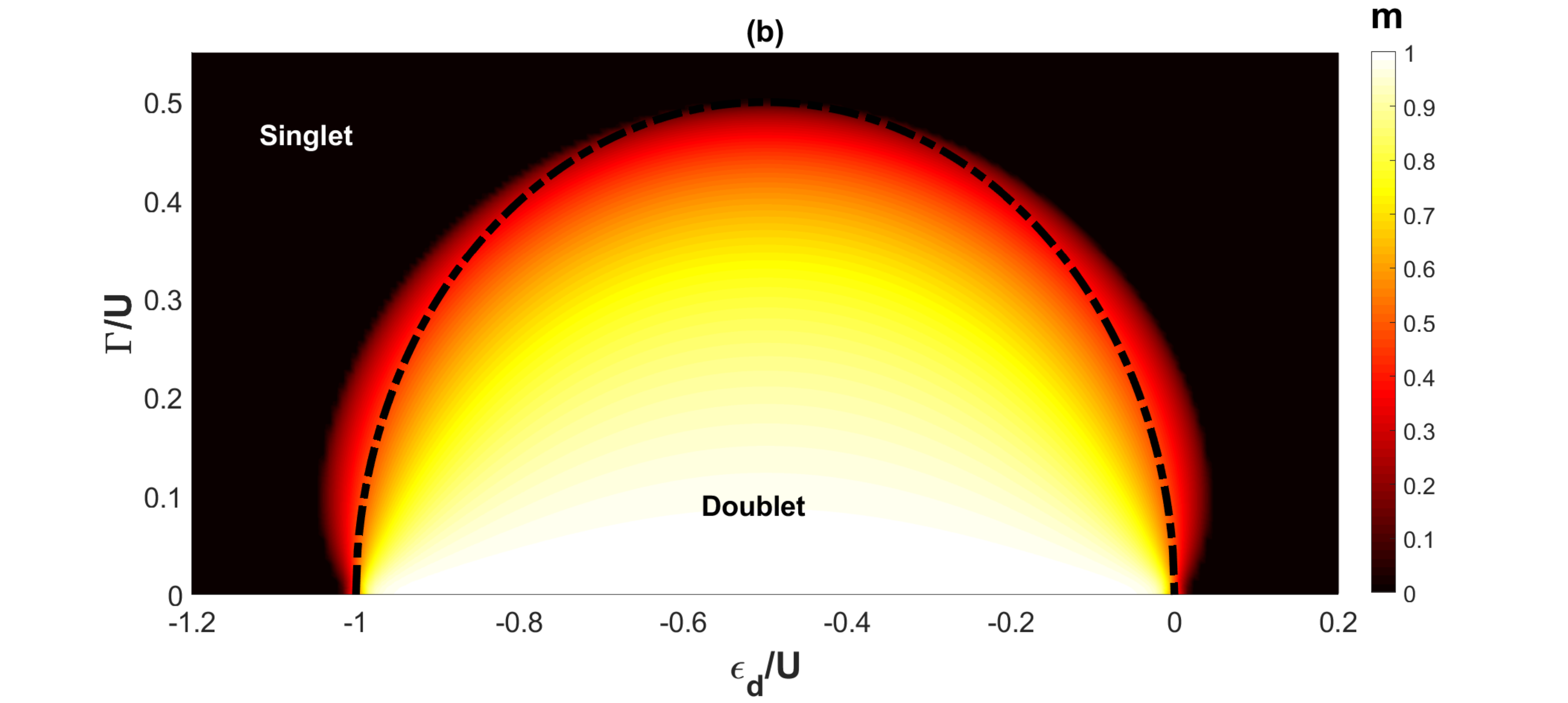}
  \centering
  \caption{Ground state phase diagram showing the  non-magnetic(Singlet)and  magnetic(Doublet) regions for quantum impurity embedded in the metal in  a) normal state ($\Gamma_N$ is the coupling between impurity and normal metallic lead) and b) superconducting state for $|\omega|<<\Delta_{sc}$.The black dashed line shows the phase boundary obtained from effective Hamiltonian\cite{Bauer2007}.}
  \label{fig:nonfloat}
\end{figure}\\
For finite superconducting gap the self-consistent treatment of $\langle n_{d\uparrow}\rangle$, $\langle n_{d\downarrow}\rangle$ and  $\langle{d_{\uparrow} d_{\downarrow}}\rangle$ (Eqs.(36) and (37)) is necessary to study the ABS and BCS singlet to magnetic doublet transition.\\
\begin{figure}[h]
  \includegraphics[scale=0.255]{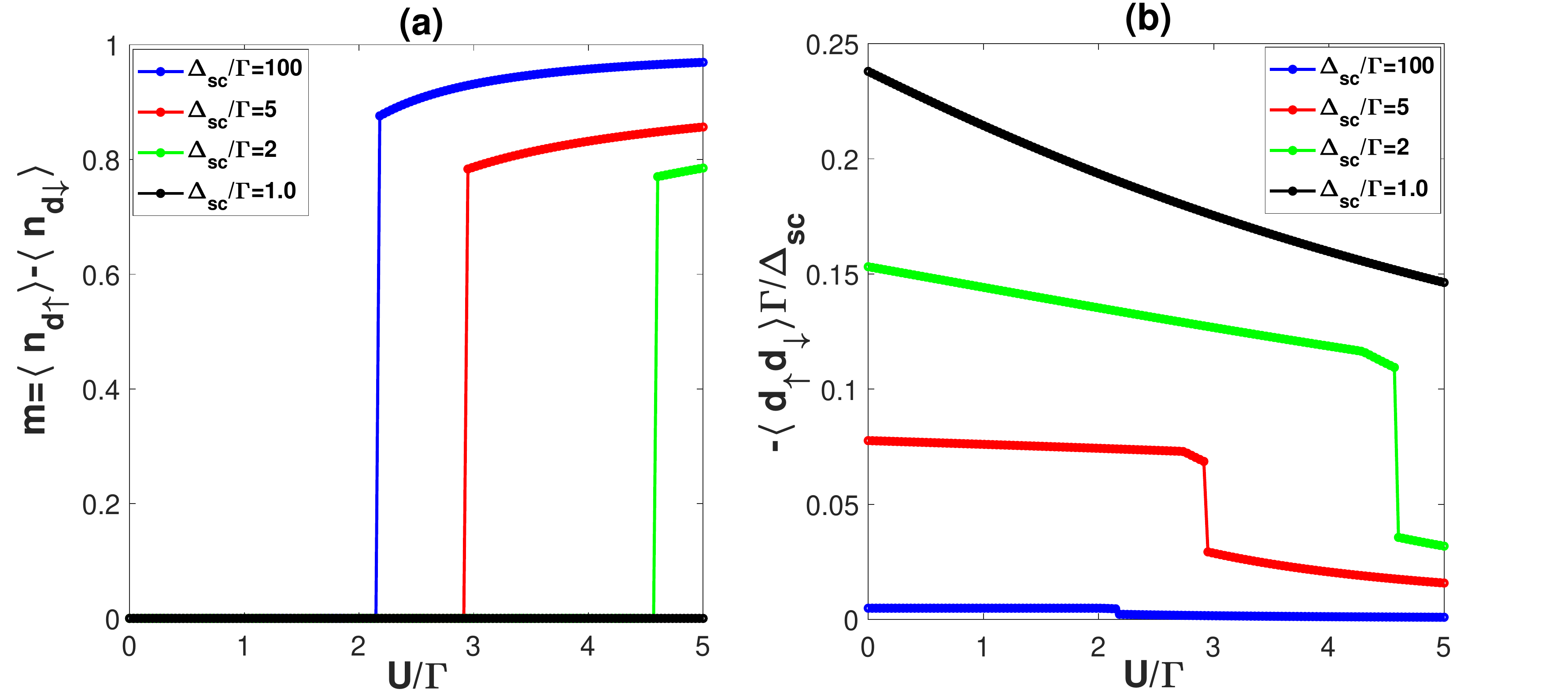}
  \centering
  \caption{$U/\Gamma$ dependence of the {\bf{(a)}} magnetic moment $m$ and {\bf{(b)}} scaled paring parameter  $\langle{d_{\uparrow} d_{\downarrow}}\rangle$ for different values of $\Delta_{sc}/\Gamma$ at the impurity site embedded in S.C for $\epsilon_d=-U/2$ at $T =0K$.}
  \label{fig:nonfloat}
\end{figure}
  Fig.6(a) and fig.6(b) shows the dependence of magnetic moment $m$ and paring parameter $\langle{d_{\uparrow} d_{\downarrow}}\rangle$ on the Coulomb interaction $U$ for various $\Delta_{sc}$. The discontinuous jump in the magnetic moment at singlet to doublet transition is different from N-QD \cite{Anderson1961} and S-QD in $|\omega|<<\Delta_{sc}$ limit and from Yoshioka \cite{Yoshioka2000} which shows that magnetic moment increases continuous from zero.  Also, the paring parameter decreases for increasing the on-site Coulomb interaction due to the suppression of superconducting correlation by repulsive coulomb interaction and become discontinuous at singlet to doublet transition. The ground state is always singlet for $U/\Gamma<2$ and it can become doublet when $U/\Gamma$ is increased. The value of Coulomb interaction $U$ at which the transition occurs decreases with increasing the value of superconducting gap $\Delta_{sc}$. This gives the indication that, the magnetic doublet region is enhanced by increasing the superconducting gap. For a large superconducting gap, i.e., $\Delta_{sc}=100\Gamma$ the non-magnetic to magnetic transition occurs close to $U/\Gamma=2$ and the induced gap become equal to $\Gamma$ in singlet state and zero in doublet similar to the low frequency or superconducting atomic limit for electron-hole symmetric case
      \begin{figure}[h]
  \includegraphics[scale=0.44]{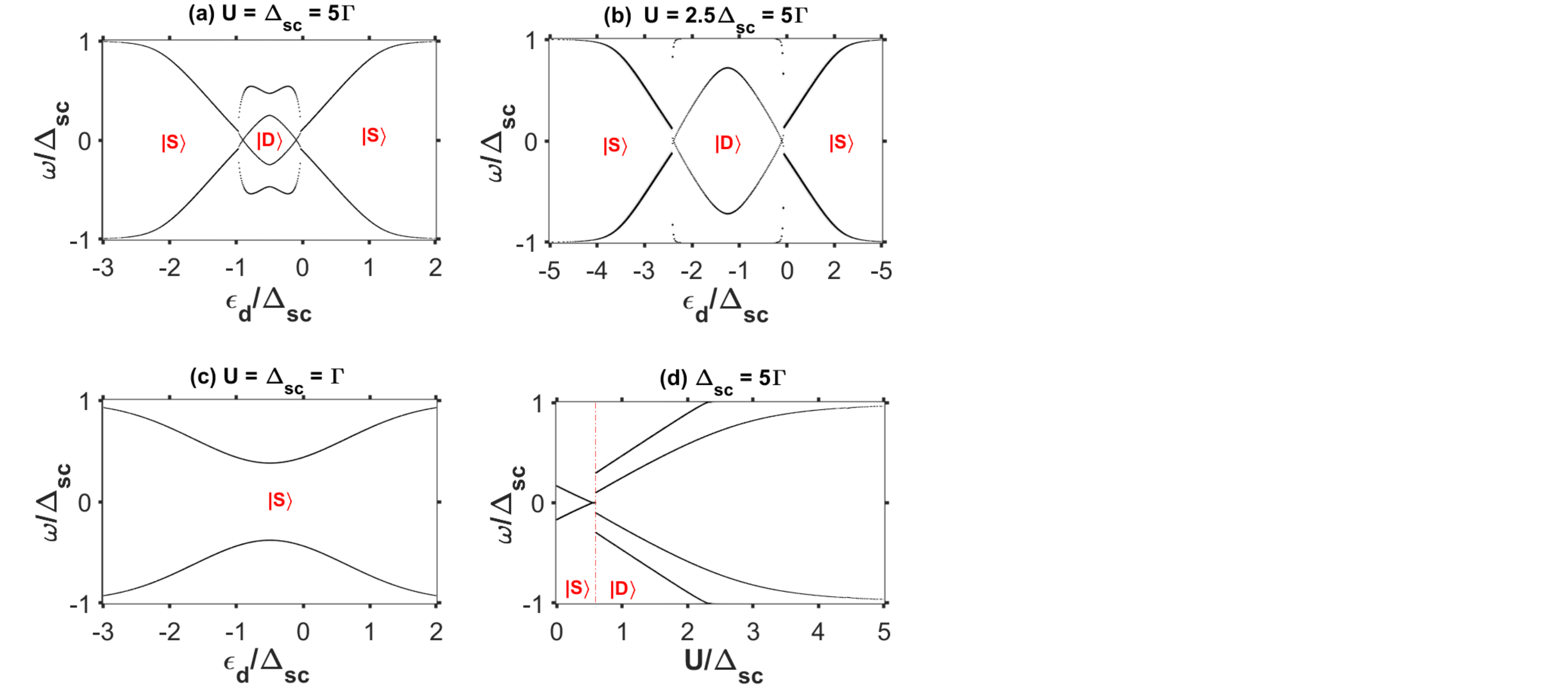}
  \centering
  \caption{Subgap ABS as a function of dot energy level $\epsilon_d/\Delta_{sc}$ (a,b,c) and  Coulomb interaction $U/\Delta_{sc}$ (d) for different $\Delta_{sc}$ values in the electron-hole symmetric case.(This is a colormap in which white and black color corresponds to $\rho_d(\omega)=0$ and $\rho_d(\omega)=1$ respectively)}
  \label{fig:nonfloat}
\end{figure}\\
To provide a complete picture of S-QD within HFA, we plot the Andreev bound states as a function of dot parameters. In fig.8(a),(b) and (c), we plot the ABS as a function of $\epsilon_d/\Delta_{sc}$ for different $U/\Delta_{sc}$ and $\Delta_{sc}/\Gamma$ ratio.  The number of Andreev bound states depends on the ratio $U/\Delta_{sc}$ ,$\Delta_{sc}/\Gamma$ and $\epsilon_d/\Delta_{sc}$. In the singlet region, only two ABS appears symmetrically with respect to the Fermi level, whereas in the doublet case, the number of ABS is doubled. Fig.8(d) shows the ABS as a function of  $U/\Delta_{sc}$ for $\Delta_{sc}=5\Gamma$. It is also clear that the outer ABS merges with the gap edge for a larger value of $U/\Delta_{sc}$. These results qualitatively agree with the recent experimental study of superconductor quantum dot nanostructures \cite{Lee2014,Pillet2013}\\
  In conclusion, to gain insights into the physics of hybrid superconductor-quantum dot devices, we considered the uncorrelated and correlated quantum impurity embedded in the BCS superconductor host. For correlated quantum impurity ($U>0$), we analyzed the low-frequency limit ($|\omega|<<\Delta_{sc}$) and finite $\Delta_{sc}$ case within the HFA. Our study for the weak coupling regime ($\Gamma<<\Delta_{sc}$ and $\Gamma\leq\Delta_{sc}$) can be regarded as complementary to the previous numerical renormalization group analysis of S-QD by J. Bauer et al \cite{Bauer2007} in the strong coupling regime ($\Gamma>>\Delta_{sc}$). The major difference between the two regimes is the nature of singlet. In the weak coupling regime, the singlet ground state corresponds to an s-wave pair, i.e., BCS singlet, whereas in the strong coupling regime the screened local spin, i.e., a Kondo singlet competes with the BCS singlet (more precisely the superconducting gap ($\Delta_{sc}$) competes with the Kondo temperature ($T_K$)).\\
  The low-frequency limit is difficult to realize exactly in the experiments, but it indicates the necessary condition for the formation of a magnetic moment at the impurity site and allows one to study sub-gap states to an extent. The competition between the proximity induced local pairing $\Delta_d$ and Coulomb interaction  $U$ on the dot site results in a transition from the BCS like the state to the singly occupied one. It is clear from our analysis that the effective local Hamiltonian phase diagram \cite{Bauer2007,Meng2009} is recovered by the Hartree Fock treatment of the complete S-QD Hamiltonian in the low-frequency limit. For the finite superconducting case, we study the singlet doublet transition and the proximity induced pairing parameter as a function of $U/\Gamma$ for different superconducting gap $\Delta_{sc}$ in the electron-hole symmetric case. We also analyze the quantum dot's spectral density to study the ABSs as a function of dot parameters. We found that superconductivity assists the formation of the magnetic moment and the doublet magnetic region is maximum for $\Delta_{sc}>>\Gamma$, which further reduces with decreasing the value of the superconducting gap and even become smaller than the magnetic doublet phase in N-QD. However, for the strong coupling regime with strong on-site Coulomb interaction, $U$, and below the Kondo temperature ($T_K$), the local magnetic moment at the impurity site can be screen by the conduction electrons at the Fermi energy (Kondo effect), and thus competes with the superconducting gap ($\Delta_{sc}$) for the singlet-doublet transition \cite{Bauer2007,Maurand2012,Zitko2015}. Our mean-field analysis provides a basis to study such strongly correlated case for one-to-one quantitative comparison between experiment and theoretical results.
Further, the above Hartree-Fock mean-field analysis can be extended to study the magnetic, spectral, and transport properties of multi-dot and multi-terminal superconductor-quantum dot devices.
\begin{acknowledgements}
One of the authors, Sachin Verma, is presently a research scholar at the department of physics IIT Roorkee and is highly thankful to the Ministry of Human Resource Development (MHRD), India, for their financial support, in the form of Ph.D. fellowship.
\end{acknowledgements}

\end{document}